\documentclass[prd,amsmath,notitlepage,twocolumn,nofootinbib]{revtex4-1}
\usepackage{graphicx}
\usepackage{amsmath}
\usepackage{amssymb}
\usepackage{mathrsfs}
\usepackage{slashed,wrapfig}
\usepackage{tikz}
\usetikzlibrary{arrows,calc,decorations.markings,math,arrows.meta}
\usetikzlibrary{shapes}
\tikzset{My Arrow Style/.style={single arrow, fill=orange!50, anchor=base, align=center,text width=2.8cm}}

\newcommand\beq{\begin{equation}}
\newcommand\eeq{\end{equation}}
\newcommand{\bea}{\begin{eqnarray}}
\newcommand{\eea}{\end{eqnarray}\noindent}

\newcommand{\Tr}{\text{Tr}}
\graphicspath{{fig/}}

\begin{document}
\title{Symmetry breaking and lattice kirigami: finite temperature effects}

\author{Antonino Flachi}
\email{flachi@keio.jp}
\author{Vincenzo Vitagliano}
\email{vincenzo.vitagliano@keio.jp}
\affiliation{Department of Physics \& Research and Education Center for Natural Sciences, Keio University, 4-1-1 Hiyoshi, Kanagawa 223-8521, Japan}
\begin{abstract}{

Recent work has analysed how deformations due to the insertion of a defect in a flat hexagonal lattice affect the ground state structure of an interacting fermion field theory. Such modifications result in an increase of the order parameter in the vicinity of the defect and can be explained by a kirigami effect, that is the combined effect of the curvature, locally introduced by the deformation in the lattice tessellation, and of a synthetic gauge field induced by the boundary conditions along the cut, performed to introduce the defect. In this work, we extend the formalism and previous results to include finite temperature effects.
}\end{abstract}
\keywords{quantum fields in curved space; chiral fermions; zeta function regularization; graphene; strongly correlated electrons}
\maketitle

%\received{} 		%%
%\revised{}
%\accepted{}		%% These are for published papers.
\section{Introduction}

Quasiparticles in quantum materials are influenced by the configuration of the crystals in which they move. In realistic crystals, in fact, the unavoidable presence of defects induces an effective change in the topology and the geometry of the lattice, with drastic consequences on quasiparticles propagation. The rapid development of this new subject in the context of quantum fields in condensed matter systems has sparked the interest in a rejuvenation of methods and results of semiclassical gravity. In the continuum limit, one can think of the lattice in which these particles move as a curved background, and apply some of the well-established techniques of quantum field theory in curved space \cite{PandT,BandD}.

Most of all, graphene \cite{CastroNeto:2009zz} is the example that recently has attracted attention on both the theoretical and applicative grounds. Graphene is the archetype of the quantum material realising an emergent system of relativistic fermions moving in a bidimensional space. 
Since its first synthesis in 2004 \cite{novo1,novo2} (but even before, following some pioneering experiment in the 60s \cite{preh,preh2,preh3}), much work has been devoted to the methodic engineering of those conditions that alter the conductivity properties of graphene. Fermions self-interactions \cite{Kotov:2010yh,Gonzalez:2000ovj}, impurities arising in the fabrication process \cite{defects,defects2}, the substrate on which the monolayer graphene sample lies \cite{wangqh}, all of them could in principle be responsible for the appearance of a mass gap in the energy spectrum of the quasiparticles propagating on the carbon sheet. The case of crystal defects is particularly interesting: several kinds of structural defects have been isolated and investigated since early experiments, among the others Stone-Wales defects (a pair of pentagons, separated by a pair of heptagons) \cite{SW}, mitosis defects (a pair of heptagons, separated by a pair of pentagons) \cite{zsoldos}, single and multiple vacancies, line defects (chains of pentagons, heptagons and hexagons gluing two patches of the crystal) \cite{line}, or out of plane carbon adatoms. Quite surprisingly, the competition between the topological features and curvature arising after the introduction of the defects has shown definitely non-trivial aspects \cite{Castro:2018iqt}.

Similar considerations have been raised for more complex crystalline structures: for example Weyl semimetals \cite{2016NatCo...711615A,2015NatPh..11..748X,2015NatPh..11..728Y,2015NatPh..11..724L,2015Sci...349..613X,2015NatPh..11..645S,2015PhRvX...5c1013L,2015NatCo...6E7373H,2015PhRvX...5a1029W,2011PhRvB..83t5101W}, three-dimensional topological materials whose low energy excitations are Weyl particles, that is massless Dirac particles with fixed chirality. Novel developments have enlightened the response of these materials to geometrical deformation and/or electromagnetic stimulation. In such a context, classical Einstein-Cartan geometry (see \cite{luz} and reference therein) provides the best framework within which modelling the effective theories that describe exotic quantum phases, with torsion-induced defects contributing in a critical way to current anomalies \cite{2017PhRvB..95k5410C,2016PhRvB..94s5112Y,2016PhRvB..94h5102Y,2016PhRvL.116p6601S,2014PhRvD..90j5004P,2013PhRvD..88b5040H,Ferreiros:2018udw}.

Fermion conductivity of quantum materials is in general sensible not only to the intrinsic conformational properties of the hosting lattices but also to the temperature of the system and eventually to the occurrence of inhomogeneous and anisotropic phases. Electrical conductivity in graphene has been the object of investigation in the framework of the Dirac model, at arbitrary values of temperature and chemical potential, in both the gapless and gapped configurations (see, e.g., \cite{Klim}). Here we are interested in studying the interplay between the effects due to finite temperature and finite density with those related to geometrical issues on a system of %strongly 
interacting particles lying on a hexagonal (graphene-like) lattice. 

As previously mentioned, four-fermion interactions are amongst the possible concause for the emergence of a non-vanishing mass gap in graphene: collective excitations arise as a consequence of some symmetry breaking mechanism (for graphene this could happen for example for breaking both $\mathbb{Z}_2$ and the discrete sublattice symmetry, which leads to staggered magnetisation). However, the occurrence of symmetry breaking in a nonflat manifold is modified by the presence of curvature \cite{dynsymbre}, with the Ricci scalar entering the gap expression in terms of an effective mass, thus contributing to the total mass of the condensate. The problem that this paper will address can be formulated as how the gap shift determined by geometry affects the phase diagram of the interacting field theory on the honeycomb lattice. Previous results on QCD phase transitions in a strong gravity environment \cite{ninoetanaka,ninosolo} have tracked the route for the development of a formalism based on the use of the effective action and zeta function regularisation to study particle condensation under the influence of external fields in a non-perturbative way. In the next sections, we will see how this proposal can be extended to the case of symmetry breaking on engineered curved lattices. 

\section{The %Hubbard 
model}

A simple, while still effective, description of strongly correlated particles on a lattice is given by the single-band Hubbard model \cite{scalettar}: being $u^\dagger, u$ and $n$ respectively creation, annihilation and number operators, the Hubbard Hamiltonian reads
\begin{align}\label{hubbard}
    H=-t\sum_{<\bf{j},\bf{l}>,\sigma}\left(u^{\dagger}_{\bf{j},\sigma}u_{\bf{l},\sigma}+\textrm{Herm. conj.}\right)+\nonumber\\
    +U\sum_{\bf{j}}n_{\bf{j}\uparrow}n_{\bf{j}\downarrow}-\mu\sum_{\bf{j}}(n_{\bf{j}\uparrow}+n_{\bf{j}\downarrow})\,;
\end{align}
the first term expresses the kinetic energy of the system: a particle with spin projection $\sigma\,(=\,\uparrow,\downarrow)$ is destroyed in the $\bf{l}$-site and created in the $\bf{j}$-site; in the Hubbard approximation, this particles' {\em hopping} is allowed only between two adjacent sites (this is the meaning of the symbol $<\bf{j},\bf{l}>$), corresponding to the positions of two first neighbour atoms on the hexagonal lattice; this assumption is justified by the exponential drop-off of the particles' wavefunctions profiles, with the hopping energy scale $t$ set by the overlap of two contiguous wavefunctions. The second term in \eqref{hubbard} controls the Coulombian repulsive interaction energy between two particles with opposite spin occupying the same site $\bf{j}$, with a coupling strength set by the parameter $U>0$. Finally, the third term, introduced by a non-vanishing chemical potential $\mu$, describes the particles filling of the lattice sites. 

The Hubbard Hamiltonian is invariant under a global SU(2) spin transformation and a U(1) (charge) redefinition of the one-particle wavefunction. On top of these, further symmetries arise if the space where the particles live is a lattice with a {\em bipartite} structure. 
%\begin{wrapfigure}{r}{4cm}
\begin{figure}[b!]
\includegraphics[width=5cm,angle=90,origin=c]{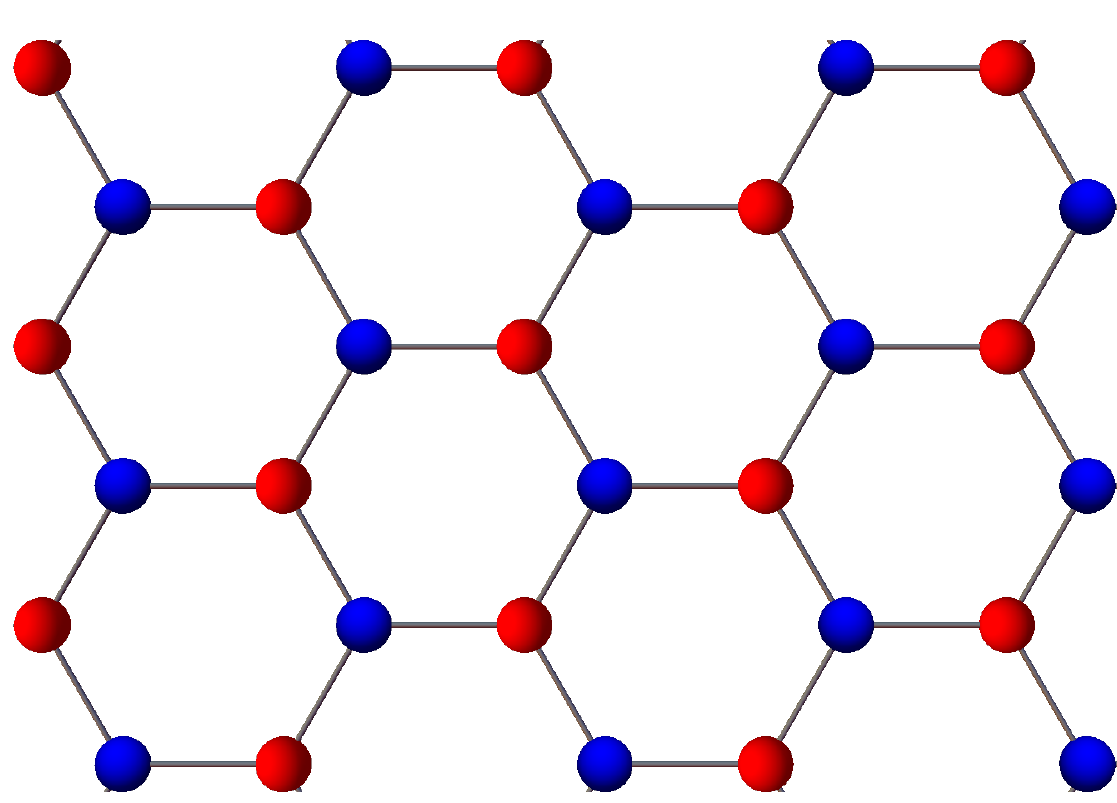}
\caption{The hexagonal honeycomb lattice with the two triangular sublattices (blue and red spheres).}\label{palline}
\end{figure}
%\end{wrapfigure}
A bipartite lattice is obtained from the union of two interpenetrating sublattices $\mathcal{A}$ and $\mathcal{B}$ (e.g. the red and blue dots triangular sublattices of Fig. \ref{palline}) such that the first neighbours of an $\mathcal{A}$ site are all $\mathcal{B}$ atoms. Hexagonal and square lattices are bipartite lattices. Triangular lattices are not. Bipartite lattices' energy is trivially minimised by N{\'e}el states, where the spins of one sublattice are all parallel among them and antiparallel to the spins of the second sublattice. In lattices which are not bipartite, instead, the N{\'e}el state is generally degenerate with other classical configuration obtained from the former by a local spin flip; these systems are said to be {\em geometrically frustrated}: half-filled lattice configurations, i.e. systems for which each site in the lattice has one and only one fermion, always have at least a pair of contiguous sites occupied by particles spinning in the same direction.

When the background on which the particles move is a bipartite lattice, the Hubbard Hamiltonian enjoys additional symmetries: it is invariant under a particle-hole transformation, that is under the change of creation into annihilation operators (and vice versa), $v^{\dagger}_{\mathbf{j},\sigma}=(-1)^{\mathbf{j}}u_{\mathbf{j},\sigma}$, such that $v^{\dagger}_{\mathbf{j},\sigma}v_{\mathbf{j},\sigma}=1-u^{\dagger}_{\mathbf{j},\sigma}u_{\mathbf{j},\sigma}$. Moreover, its spectrum remains unchanged under a sign flip of the hopping parameter $t$.

The influence of temperature and geometry on the occurrence of symmetry breaking is more conveniently studied in a covariant functional-integral formulation of the Hubbard model associated partition function, which can be straightforwardly achieved \cite{path, path2, herbut} bosonising the density-density product in the interaction term of \eqref{hubbard}. An explicitly SU(2)-invariant Hubbard-Stratonovich transformation of the interaction term may be obtained, for spin-$1/2$ fermions, by introducing an arbitrary unit vector $\mathbf{n}$ along the spin-quantization axis of the particles, such that
\begin{align}
n_{\bf{j}\uparrow}n_{\bf{j}\downarrow}=\frac{1}{4}(n_{\bf{j}\uparrow}+n_{\bf{j}\downarrow})^2-(\mathbf{S} \cdot \mathbf{n} )^2\nonumber\\
 \mathbf{S}=\frac{1}{2}\sum_{\sigma,\tau} u^{\dagger}_{\mathbf{j},\sigma}\vec{\Sigma}_{\sigma\tau} u_{\mathbf{j},\tau}\,,
\end{align}
where $\vec{\Sigma}=(\sigma^x,\sigma^y,\sigma^z)$ is the usual Pauli vector. Introducing two auxiliary fields $\phi$ and $\Delta$, the density-density operator can be rewritten as
\begin{align}
&e^{U\sum_j n_{\mathbf{j}\uparrow}n_{\mathbf{j}\downarrow}}=\nonumber\\
&=\int\prod_j\frac{d \phi_\mathbf{j}\,d\Delta_\mathbf{j}\,d^2\mathbf{n}_\mathbf{j}}{4\pi^2 U}\,e^{\left[\frac{\phi_\mathbf{j}^2}{U}+\imath\phi_\mathbf{j}(n_{\mathbf{j}\uparrow}+n_{\mathbf{j}\downarrow})+\frac{\Delta_\mathbf{j}^2}{U}-2\Delta_\mathbf{j}\,\mathbf{n}_\mathbf{j}\cdot\mathbf{S}_\mathbf{j}\right]}\,.
\end{align}
%\begin{widetext}
%\begin{align}
%\textrm{Exp}\,\left[U\sum_j n_{\mathbf{j}\uparrow}n_{\mathbf{j}\downarrow}\right]=\int\prod_j\frac{d \phi_\mathbf{j}\,d\Delta_\mathbf{j}\,d^2\mathbf{n}_\mathbf{j}}{4\pi^2 U}\textrm{Exp}\left[\frac{\phi_\mathbf{j}^2}{U}+\imath\phi_\mathbf{j}(n_{\mathbf{j}\uparrow}+n_{\mathbf{j}\downarrow})+\frac{\Delta_\mathbf{j}^2}{U}-2\Delta_\mathbf{j}\,\mathbf{n}_\mathbf{j}\cdot\mathbf{S}_\mathbf{j}\right]\,.
%\end{align}
%\end{widetext}
The effective Lagrangian for the fluctuations is then derived integrating out the fermionic degrees of freedom. 

The procedure here described is entirely general; however, to ease the calculation, we will consider only one scalar order parameter, $\phi_\mathbf{j}$, breaking at the same time the $\mathbb{Z}_2$ and the discrete sub-lattice symmetry of the half-filled Hubbard model on a hexagonal lattice. This order parameter can be identified with the staggered magnetisation -- the conjugated of the density operator -- defined as the net difference between the number of negatively and positively spinning fermions, $\phi_{\mathbf{j}}\equiv n_{\mathbf{j}\uparrow}- n_{\mathbf{j}\downarrow}$. 

For sake of compactness, it is possible to introduce a new wave function $\psi_\sigma$ such that $\psi_\sigma^T\equiv(\psi_\sigma^{\mathcal{A}1},\psi_\sigma^{\mathcal{B}1},\psi_\sigma^{\mathcal{A}2},\psi_\sigma^{\mathcal{B}2})$, where $X=\mathcal{A}$ and $\mathcal{B}$ refer to the two sublattices, $\#=1$ and $2$ are the two inequivalent Fermi points $K^\#$ of the half-filled honeycomb lattice, and $\psi^{X,\#}_\sigma$ are the inverse Fourier transforms of the sublattice annihilation operators $u_{X,\sigma}(\mathbf{p}+K^\#)$. With these assumptions the effective Lagrangian finally reads
\begin{align}\label{masterL}
    \mathcal{L}=\overline{\psi_\sigma}(\imath\gamma^a\partial_a+\mu\gamma^0){\psi}_\sigma+\alpha_\sigma\overline{\psi_\sigma}\phi\,{\psi}_\sigma+\frac{\phi^2}{2 \lambda}\,,
\end{align}
expressed in terms of the usual flat space $\gamma$-matrices, $\gamma^a$;  here $\lambda$ is a constant, proportional to the coupling strength parameter $U$ up to an irrelevant factor, and $\alpha_{\uparrow,\downarrow}=\pm1$.

\section{{\em Kirigami} %art 
and lattice geometry}

\begin{figure}[bp!]
 \includegraphics[width=\linewidth]{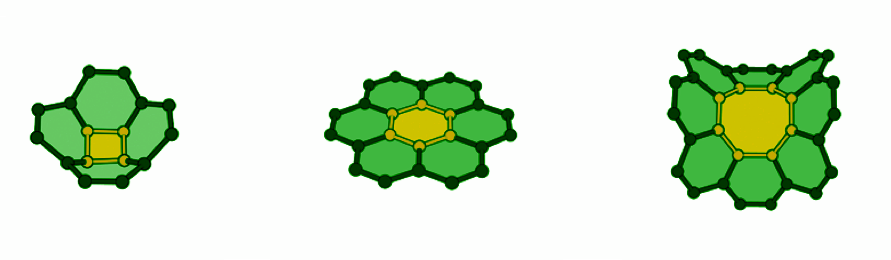}
\caption{The flat space is modified through the insertion or the extraction of a piece of lattice. Adding a section with a given angle, one finds a saddle; subtracting the same section, and then sewing on the cut, a cone. CCDC codes: [4]-circulene $\Rightarrow$ (747755), [6]-circulene $\Rightarrow\!$ (1129883),
[8]-circulene $\Rightarrow\!$ (1106253)}\label{graphene}
\end{figure}

The Japanese term {\em kirigami} describes a particular origami technique which, apart for folding the paper, allows for cutting parts of it. Two-dimensional carbon allotropes have recently been shown to be perfect candidates for engineering robust three-dimensional microscale structures with tunable mechanical properties (see for example \cite{kirigami}), making of them a sort of ``lattice kirigami''. The most straightforward procedure to obtain such kirigami is typically via defect insertion in the (honeycomb, in the present case) lattice, as shown in Fig. \ref{graphene}. 
Starting from a locally flat lattice where a hexagonal cell is marked as the centre, one can design two kinds of defects, bringing to two different geometries: reducing the number of the central cell sides of $N>0$ sides corresponds to cut and pulling out a $N \pi/3$ section of the lattice; gluing together the two sides results then in a conical configuration with positive curvature. On the other side, augmenting the number of sides (which in our notation corresponds to consider a negative $N$) of the central cell is equivalent to pull into the lattice a $|N|\pi/3$ sector, resulting in a saddle geometry with negative curvature. We will only consider lattices with an even number of defects, preventing possible frustration effects from veiling or even obscuring the main quantum mechanisms here studied.

The kirigami procedure, as stated above, alters the topological and geometrical features of the material sheet: the very same fermions propagation is sensitive to these conformational properties of the underlying lattice. As a result, the study of the vacuum structure and the symmetry breaking behaviour exhibited by a continuous field theory like  \eqref{masterL} defined on the curved lattice, relies on the description of an effective continuous geometry of the lattice itself.

In the continuum limit, %it is natural to assume 
this configuration of the lattice can be described by a conical metric  \cite{somm,fur1,fur2}. Away from the central defect (corresponding to a conical singularity), the associated spacetime is accurately described by a locally flat geometry, $ds_{\textrm{con}}^2=dt^2-dr^2-r^2d\widetilde{\theta}^2$, but with an important {\em caveat}: it is not globally Euclidean, since the angular coordinate does not run on a $2\pi$ circle; instead, $0\leq\widetilde{\theta}<2\pi -\Delta$, with $\Delta>0 \;(\Delta<0)$ being the deficit (excess) angle: surfaces at constant $t$ are cones (saddles), not planes. It is worth to stress that in the case under examination (lattice with a hexagonal base) $\Delta$ does not take continuous values: it is clear that it should be $\Delta=N\pi/3$, with $|N|$ even to avoid the frustration of the system mentioned above. Note also that there is a constitutive bound on $N$: while the only positive value allowed is $N=2$ (corresponding to the square defect), negative values are allowed up to $|N|\sim16$, on top of which the geometrical structure is supposed to collapse onto a helical conformation \cite{rickhaus}. 

%\begin{wrapfigure}{l}{6cm}
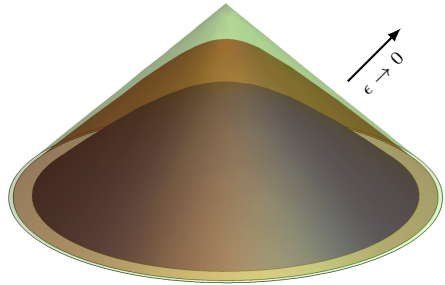
\begin{figure}
\includegraphics[width=6cm]{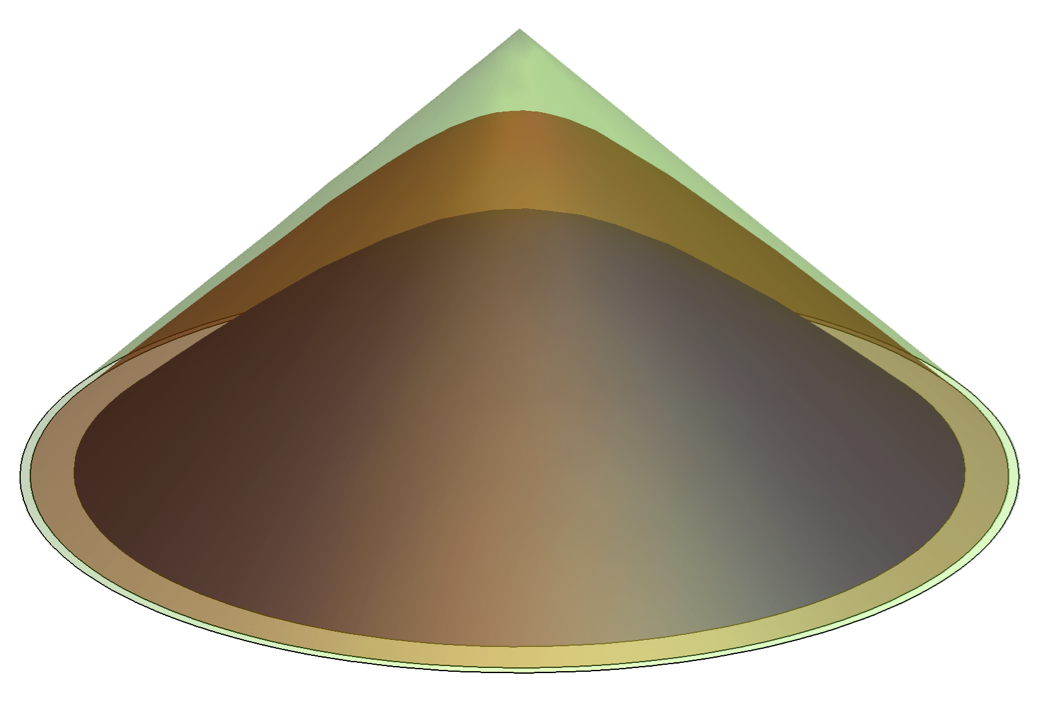}
  \put(-40,80){\rotatebox{45}{\begin{tikzpicture}[thick]  
\draw[-{latex[scale=3.0]}] (0,0) -- (1,0);
\end{tikzpicture}}}
\put(-35,75){\rotatebox{45}{{\scriptsize $\epsilon\rightarrow0$}}}
\caption{A sequence of regular surfaces with smoothed tips corresponding to the spatial part of \eqref{sequence}, with $\alpha=2/3$ ($\Delta=2\pi/3$). The sequence approximates the conical metric, where $\epsilon$ vanishes identically.}\label{coni}
\end{figure}
%\end{wrapfigure}

The conical metric, $ds_{\textrm{con}}^2$, is singular on the tip, where it is not possible to introduce a tangent space and consequently, is not possible to calculate the curvature. This singularity is only an artefact of our mathematical description of the cone: the deformed lattice %kirigami itself 
does not present any singularity in correspondence of the defect. A better approximation of the physical lattice is achieved smoothing the geometry into that of a snub-nosed cone (resp. saddle), with metric $ds_\epsilon^2$ regulated by a parameter $\epsilon$. The biggest is the regulator $\epsilon$, the smoothest the tip of the cone. The proper conical metric $ds_{\textrm{con}}^2$ is recovered as the limiting case of a sequence of such smoothed cones, when $ds_{\textrm{con}}^2=\lim_{\epsilon\rightarrow0} ds_\epsilon^2$ (see Fig. \ref{coni}). In the lattice language, the regulator $\epsilon$ is never vanishing, rather it parametrises the minimum physical distance between adjacent atoms.

Let us introduce a new angular coordinate $\theta=\widetilde{\theta}/\alpha$, with $\alpha=1-\Delta/2\pi$, such that $0\leq{\theta}<2\pi$. The line element of the regularised tip can then be written as 
\begin{align}
    ds_\epsilon^2=dt^2-f(r,\epsilon)dr^2-\alpha^2 r^2 d\theta^2\,,
\end{align}
where the smooth function $f(r,\epsilon)$ satisfies the following asymptotics: i) $\lim_{r\rightarrow0}f(r,\epsilon)=\alpha^2$ and ii) $\lim_{r\gg\epsilon}f(r,\epsilon)=1$. The simplest choice implementing these requirements is the sequence of spaces with metric
\begin{align}\label{sequence}
    ds_\epsilon^2=dt^2-\frac{r^2+\epsilon^2\alpha^2}{r^2+\epsilon^2}dr^2-\alpha^2 r^2 d\theta^2\,.
\end{align}

The covariant extension of the model \eqref{masterL} to the curved spaces \eqref{sequence} is obtained under the minimal coupling {\em ansatz}~: the flat metric is substituted by the metric tensor $g_{\mu\nu}$ defined  in \eqref{sequence} and such that $g^{\mu\nu}=e_a^\mu e_b^\nu\eta^{ab}$, being $e_a^\mu$ the  {\em vielbein} and $\eta^{ab}$ the Minkowski metric; the ordinary partial derivative is replaced by the covariant derivative and the flat $\gamma$-matrices by $\gamma^\mu=\gamma^a\,e_a^\mu$, which are the $\gamma$-matrices in curved space.

There is still one geometrical issue arising in the kirigami procedure. The cut-and-paste technique to induce lattice defects changes the topology of the system. One can show \cite{sitenko} that the boundary conditions satisfied by the spinors along the cut read as what is evocatively called the Moebius stripe conditions: rotation around the defect brings to a configuration where the two triangular sublattices are exchanged, while one more rotation brings the system back to the initial state,
\begin{align}\label{moebius}
    \psi(r,\theta+2\pi)= - \exp\left(-i \frac{\pi}{2}N R\right)\psi(r,\theta)\,,
\end{align}
where $R=i\left(\begin{array}{cc} 0 & \sigma^2\\ -\sigma^2 & 0 \end{array}\right)=-\gamma^5$ (last equality holds in the standard planar representation, where $\gamma^0$ is diagonal, but changes in other representations of the Clifford algebra). Note that $R$ anticommutes with all the $\gamma$-matrices. Using a gauge-like transformation mapping $\psi$ into a new field $\widetilde{\psi}$ such that
\begin{align}\label{fields}
\widetilde{\psi}=\exp\!\left(\imath \theta\frac{N}{4} R\right)\!\psi\,\Rightarrow \,\psi\equiv A(\theta)\widetilde{\psi}=\exp\!\left(- \imath\theta\frac{N}{4} R\right)\!\widetilde{\psi}\,,
\end{align}
it is possible to re-arrange the Moebius boundary condition \eqref{moebius} to resemble the flat space (antiperiodic) condition
\begin{align}
\widetilde{\psi}(r,\theta+2\pi)= -\widetilde{\psi}(r,\theta)\,.
\end{align}

The effective Lagrangian \eqref{masterL}, under the fields redefinition \eqref{fields}, acquires an extra contribution in the form of a (constant) effective gauge connection term\footnote{For a similar role of synthetic gauge fields in condensed matter physics see also \cite{VozmRev,AmorimRev,Cortijo:2011aa,deJuan:2010zz,Cortijo:2006xp}.}, $\mathcal{B}_\mu\equiv-\delta_\mu^\theta\frac{N}{4}R$, so that now the bosonised Hubbard model reads
\begin{align}\label{masterL2}
    \mathcal{L}=\overline{\widetilde{\psi}_\sigma}\imath\gamma^\mu\mathcal{D}_\mu{\widetilde{\psi}_\sigma}+\mu\,\overline{\widetilde{\psi}_\sigma}\gamma^0\,{\widetilde{\psi}}_\sigma+\alpha_\sigma\overline{\widetilde{\psi}_\sigma}\phi\,{\widetilde{\psi}}_\sigma+\frac{\phi^2}{2 \lambda}\,,
\end{align}
where $\mathcal{D}_\mu=\nabla_\mu+\imath \mathcal{B}_\mu$. In what follows, we will drop all the tilde's from the previous equation to shorten the notation, having in mind the {\em caveat} that spinor fields have already undergone a Moebius stripe rotation. 

\section{%Matsubara formalism for 
Finite temperature effects
%quantum field theories
}

The effects due to finite temperature $T$ can be straightforwardly incorporated using the Euclidean Matsubara approach \cite{matsu, landsman}, that is Wick-rotating the metric \eqref{sequence}, compactifying the (imaginary) time onto the interval $0\leq\tau<\beta=1/T$ and imposing antiperiodic boundary conditions on the fermion fields in the $\tau$ direction, $\psi(\tau=0,x)=-\psi(\tau=\beta,x)$. The boundary conditions imply that fermion fluctuations can be Fourier expanded in the Euclidean time as
\begin{align}\label{modes}
    \psi(\tau,x)=\sum_{\omega_n} e^{\imath \omega_n \tau}\psi_n(x)\,,
\end{align}
 where $\omega_n=(2n+1)\pi /\beta$ are the quantised (Matsubara) frequencies.
 
In order to understand the phase diagram of the lattice kirigami we need to study the behaviour at different temperature of the staggered magnetisation parameter $\phi$, which can be calculated solving numerically the effective equations of motion for the system.  Keeping this strategy in mind, we start by evaluating the Euclidean effective action for the model \eqref{masterL2} on the Wick-rotated curved background \eqref{sequence},
\begin{align}\label{eff1}
\Gamma=&-\!\int d^3x\sqrt{g}\left(\frac{\phi^2}{2\lambda}\right)+\!\!\sum_{\sigma=\uparrow,\downarrow}\!\ln\mbox{Det} \left(\imath \gamma^\mu\mathcal{D}_\mu+\alpha_\sigma\phi+\mu\gamma^0\right)\nonumber\\
&\equiv -\int d^3x\sqrt{g}\left(\frac{\phi^2}{2\lambda}\right)+\delta\Gamma\,.
\end{align}
As it is, this expression of the effective action does not allow to obtain directly the curvature counterterms to the one loop divergences of the Dirac operator (the second term in \eqref{eff1}); this is due to the circumstance of  $\mathscr{O}\equiv(\imath \gamma^\mu\mathcal{D}_\mu+\alpha_\sigma\phi+\mu\gamma^0)$ being a linear differential operator of the first order, rather than being of second order. A possible way out is to apply standard techniques and square the Dirac operator: the eigenvalues of the squared (Dirac) functional determinant on the Riemannian spin manifold are then related to those of the sum of the spinor Laplacian with the space curvature, through the Schr\"odinger--Lichnerowicz--Weitzenb\"ock formula,  $\slashed{\mathcal{D}}^2=\nabla^*\nabla+\mathbb{I}\cdot\mathscr{R}/4$ (see for example \cite{friedrich}). Following this prescription, the one loop effective action \eqref{eff1} is rewritten as
 \begin{align}\label{eff}
\delta\Gamma&=\frac{1}{2}\sum_{\sigma=\uparrow,\downarrow}\ln\mbox{Det} \left(\Box-\mu^2-2\imath\mu\partial_t+\mathscr{V}\right)=\nonumber\\
&=\frac{1}{2}\sum_{\sigma=\uparrow,\downarrow}\sum_{n=-\infty}^\infty\mbox{ln}\,\mbox{Det}\left(\omega^2_n-\Delta-\mu^2-2\imath\mu\omega_n\mathscr{V}\right)\,,\nonumber\\
&\hspace{2.6cm}\left(\mathscr{V}=\frac{\mathscr{R}}{4}+\phi^2+\alpha_\sigma \sqrt{g^{rr}}\phi'\right)
\end{align}
where it should be taken into account that the d'Alembertian operator, $\hat{\Box}$, is built out of the total covariant derivative including the contribution from the effective gauge connection  and calculated on the Euclidean version of the regularised metric, $ds^2_\epsilon$ (see also \cite{ninoetanaka, inagakitemp} for further details); in the last line we have taken advantage of the mode expansion \eqref{modes} in the Matsubara frequencies.
 %and have defined $\hat{\Delta}={\hat{\nabla}_j{\hat{\nabla}^j+\mathcal{B}_j}\mathcal{B}^j},\, j=\{1,2\}$. 
 The expression of the effective action \eqref{eff} can be obtained through different techniques, usually always quite cumbersome. In the present case, we will use zeta-function regularization (see for example \cite{PandT, Hawking:1976ja,buch,eliz,avra,kirs}). Given an operator $\mathscr{O}$, its functional determinant is the product of the $\mathcal{N}$ eigenvalues $\kappa_\mathcal{N}$, such that
\begin{align}\label{W}
\ln\mbox{Det}({l}^2\mathscr{O})=\sum_{\mathcal{N}}\ln(l^2\kappa_\mathcal{N})=\lim_{s\rightarrow0}\sum_{\mathcal{N}}\kappa_\mathcal{N}^{-s}\ln(l^2\kappa_\mathcal{N})\,,
\end{align}
where $l$ is some renormalization length. Defining the function 
\begin{align}
\zeta(s)=\sum_{\mathcal{N}}\kappa_\mathcal{N}^{-s}\,,\qquad \left(\zeta'(s)=-\sum_{\mathcal{N}}\kappa_\mathcal{N}^{-s}\ln\kappa_\mathcal{N}\right)
\end{align}
it is possible to rewrite the expression \eqref{W} as
\begin{align}
\ln\mbox{Det}({l}^2\mathscr{O})=\zeta(0)\ln l^2-\zeta'(0)\,,
\end{align}
provided that $\zeta$ and its first derivative are regular in zero. In three dimensions, $\zeta(s)$ converges for $\Re\,[s]>3/2$, and can be analytically continued to a meromorphic function of $s$ for $\Re\,[s]<3/2$; in particular it will be regular in zero, and will encounter simple poles for $s=(3/2-p),\, p\in \mathbb{N}$. Under these assumptions, and using the integral representation of the function $\Gamma(s)$, it is possible to relate the zeta-function to the Mellin transform of the heat trace, which in the case of the differential operator in \eqref{eff} reads
%\begin{widetext}
\begin{align}\label{zeta}
&\zeta(s)=\frac{1}{\Gamma(s)}\int^{\infty}_0d\tau \tau^{s-1}\Tr \,e^{-\tau \mathscr{O}}=\nonumber\\
&=\frac{1}{\Gamma(s)}\!\sum_{\sigma=\uparrow,\downarrow}\sum_{n=-\infty}^\infty\int^{\infty}_0\!\!\!d\tau \tau^{s-1}\Tr\, e^{-\tau \left(\omega^2_n-\Delta-\mu^2-2\imath\mu\omega_n+\mathscr{V}\right) }
\end{align} 
%\end{widetext}
In curved space, the heat kernel of a second order differential operator like the one appearing in \eqref{zeta} has a particularly enjoyable proper time expansion, exhibiting a re-summation, among the curvature invariants contributions, of all the power terms in the scalar curvature $\mathscr{R}$ \cite{Parker:1984dj, Jack:1985mw,cg1,cg2},
%\begin{widetext}
\begin{align}\label{heat}
&\Tr \,e^{-\tau\left(\omega^2_n-\Delta+\frac{\mathscr{R}}{4}+\phi^2-\mu^2-2\imath\mu\omega_n+\alpha_\sigma \sqrt{g^{rr}}\phi'\right)}=\nonumber\\
&=\frac{e^{-\tau (\omega^2_n+\frac{\mathscr{R}}{12}+\phi^2-\mu^2-2\imath\mu\omega_n+\alpha_\sigma \sqrt{g^{rr}}\phi')}}{(4\pi \tau)^{3/2}}\sum_{k=0}^{\infty} \Tr^{\textrm{\tiny{(Dirac)}}}\mathscr{D}_\sigma^{(k)}\tau^k\,,
\end{align}
%\end{widetext}
where the traces are taken on the Dirac indices and the first three $\mathscr{D}_\sigma^{(k)}$ coefficients read
\begin{gather}\label{cs}
\mathscr{D}^{(0)}=1\,,\qquad\qquad\mathscr{D}^{(1)}=0\,,\nonumber\\
\mathscr{D}^{(2)}=\frac{1}{180}\left(\mathscr{R}_{\mu\nu\rho\sigma}\mathscr{R}^{\mu\nu\rho\sigma}-\mathscr{R}_{\mu\nu}\mathscr{R}^{\mu\nu}\right)+\frac{1}{120}\Delta \mathscr{R}+\nonumber\\
\hspace{1.5cm}+\frac{1}{6}\Delta \left(\phi^2+\alpha_\sigma \sqrt{g^{rr}}\phi'\right)+\frac{1}{12}W_{\mu\nu}W^{\mu\nu}\,,
\end{gather}
with $W_{\mu\nu}=\left[\mathcal{D}_\mu,\mathcal{D}_\nu\right]=(-1/4)\mathscr{R}_{\mu\nu\rho\sigma}\gamma^\rho\gamma^\sigma$.
  
\begin{figure*}[htp!]
 \includegraphics[width=7cm]{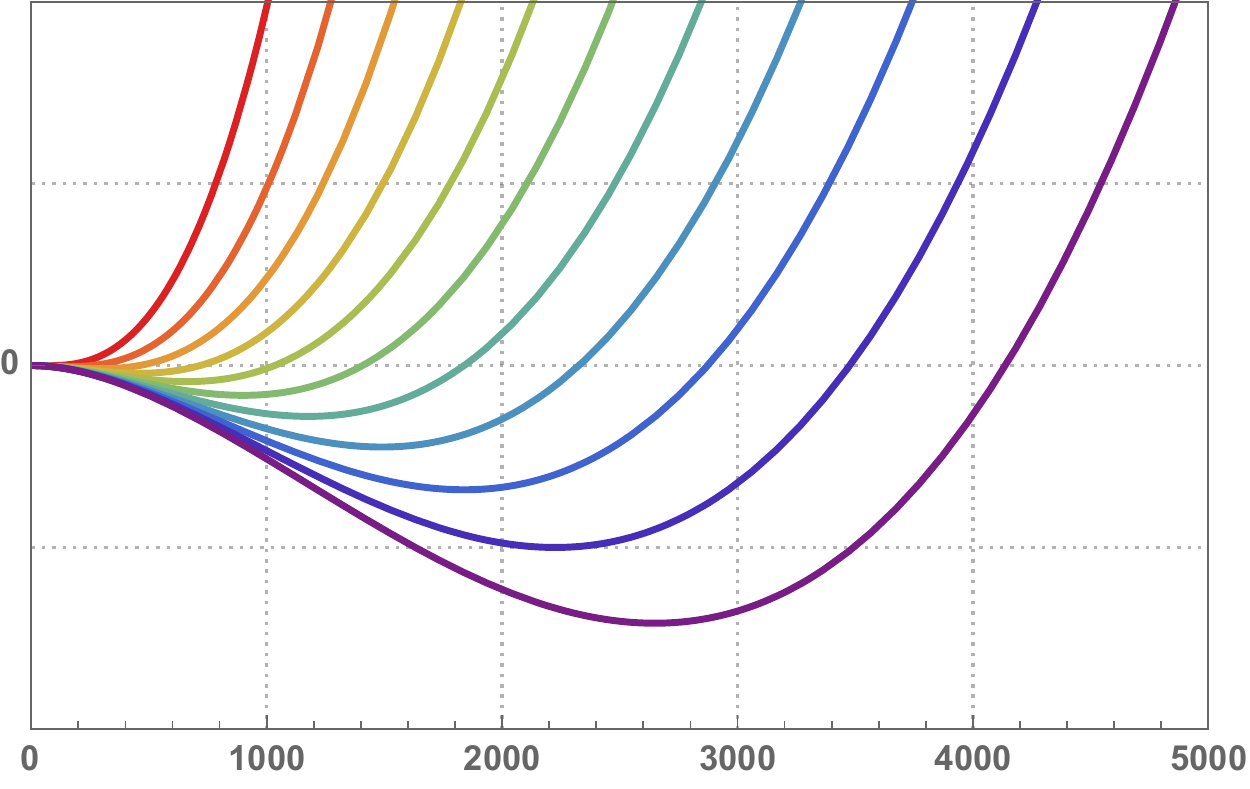}
 \hspace{0.5cm}
  \includegraphics[width=7cm]{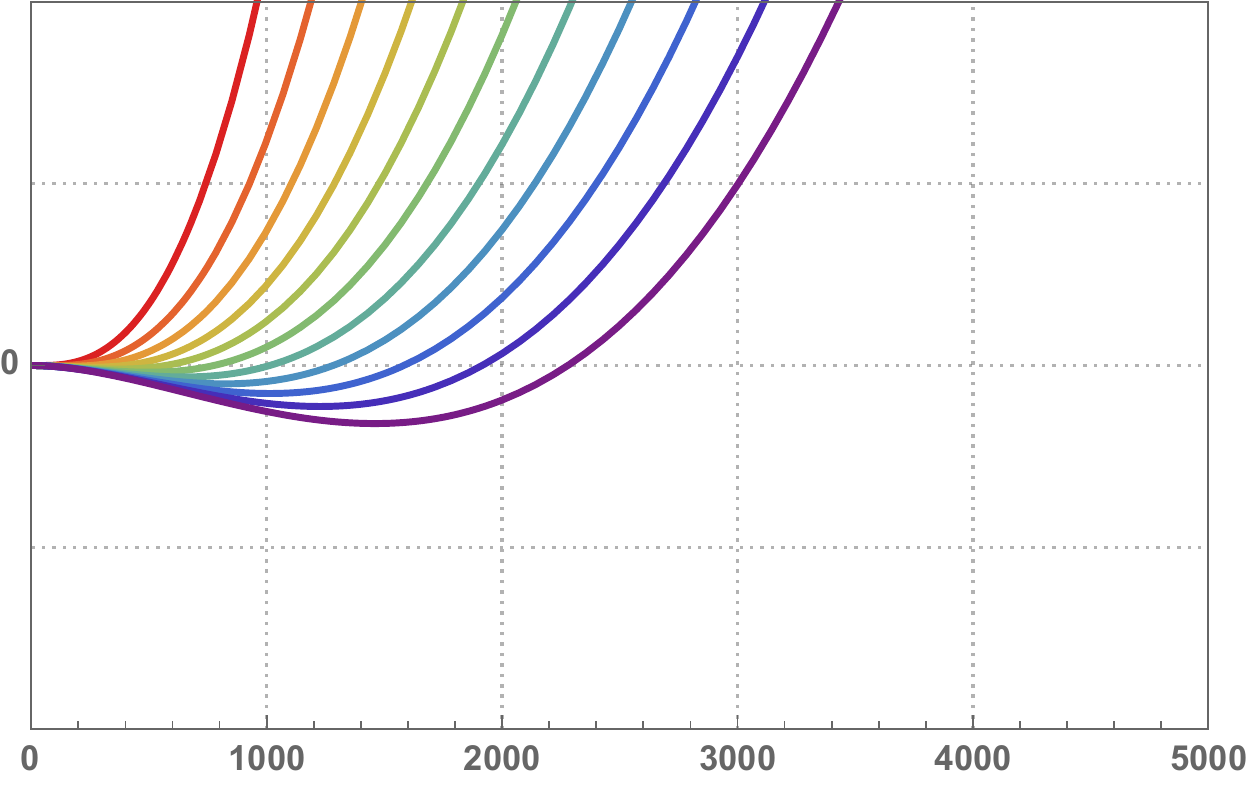}
    \includegraphics[height=4.5cm]{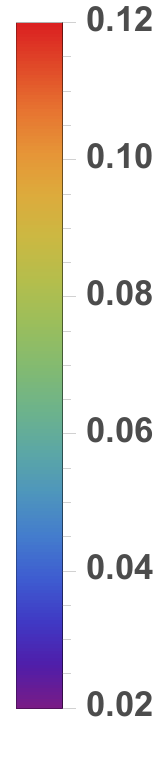}
    \put(-460,34){\rotatebox{90}{{\scriptsize Effective potential}}}
        \put(-240,34){\rotatebox{90}{{\scriptsize Effective potential}}}
    \put(-160,-8){\rotatebox{0}{{\scriptsize Order parameter, $\phi$}}}
\put(-380,-8){\rotatebox{0}{{\scriptsize Order parameter, $\phi$}}}
    \put(5,39){\rotatebox{90}{{\scriptsize Temperature, $T$}}}
\caption{The effective potential on a cone with opening angle $\alpha=2/3$, as a function of the temperature and at vanishing chemical potential. The potential is calculated very close to the defect (left panel) and at infinity (right panel). The regulating parameter is taken to be $\epsilon=0.001$.}
\label{pot}
\end{figure*}

In order to calculate the zeta-function \eqref{zeta} we will follow the strategy delineated in \cite{ninoetanaka}: the sum on the quantum number $n$ can be expressed in a more convenient way observing that (see also \cite{gusev}),
\begin{align}\label{smartsum}
%\mathscr{G}_{\beta,\mu}(\tau)\equiv
\sum_{n=-\infty}^{+\infty} e^{-\tau (\omega_n-\imath \mu)^2}=\frac{\beta}{2\sqrt{\pi\tau}}\theta_3\left[e^{-\frac{\beta^2}{4 \tau}}; \frac{\pi-\imath \beta\mu}{2}\right]=\nonumber\\
=\frac{\beta}{2\sqrt{\pi\tau}}\left[1+2\sum_{n=1}^{+\infty}(-1)^n e^{-\frac{\beta^2 n^2}{4 \tau}}\cosh(\beta\mu n)\right]\,;
\end{align}
using \eqref{smartsum} and \eqref{heat} in \eqref{zeta} one finds, after some lengthy calculation, an expression for the zeta-function. In odd dimensions, the analytic continuation of $\zeta(s)$ to $s=0$ vanishes, while the first derivative turns out to be
\begin{align}\label{zeta0}
&\zeta'(0) = \beta{1\over (4\pi)^{2}} \sum_{k=0}^\infty \sum_{\sigma}\left[a_k \mathscr{D}^{(k)}_{\sigma} X_{\sigma}^{\frac{3}{2}-k}+\right.\nonumber\\
&+\frac{\mathscr{D}^{(k)}_{\sigma}}{2^{k-\frac{5}{2}}}\left({X_{\sigma}}\right)^{\frac{3}{4}-\frac{k}{2}} 
\sum_{n=1}^\infty(-1)^n {\cosh(\beta \mu n)\over \left(n\beta\right)^{3/2-k}} K_{k-\frac{3}{2}}\!\left(n\beta\sqrt{X_{\sigma}}\right)\!\Big],
\end{align}
where $K$'s are the modified Bessel functions of the second kind (which have the nice property to be exponentially suppressed) and
\begin{gather}
a_k= \lim_{s\rightarrow 0}{\Gamma(s+k-3/2)\over \Gamma(s)}\left( \digamma\left(s+k-3/2\right) - \digamma\left(s\right)\right)
\nonumber\\
X_\sigma=\frac{\mathscr{R}}{12}+\phi^2+\alpha_\sigma \sqrt{g^{rr}}\phi'\,,
\end{gather}
being $\digamma$ the digamma function (the logarithmic derivative of the gamma function $\Gamma$). Considering only terms up to the second order in the heat kernel expansion, \eqref{zeta0} reads
\begin{align}
&\zeta'(0) = {\beta\over (4\pi)^{3/2}} \sum_{\sigma}\Bigg\{\frac{4}{3}\sqrt{\pi}X_\sigma^{3/2}+\sqrt{\pi}\mathscr{D}^{(2)}_{\sigma} X_\sigma^{-1/2}+
\nonumber\\
&+\sqrt{2}\sum_{n=1}^\infty(-1)^n \cosh(\beta \mu n)\Bigg[4 \frac{X_\sigma^{3/4}}{(n\beta)^{3/2}}K_{3/2}\left(n\beta\sqrt{X_{\sigma}}\right)+\nonumber\\
&\hspace{1.5cm}+\mathscr{D}^{(2)}_{\sigma}\frac{X_{\sigma}^{-1/4} }{(n\beta)^{-1/2}}
 K_{1/2}\left(n\beta\sqrt{X_{\sigma}}\right)\Bigg]\Bigg\}\,,
\end{align}
which, given that $\zeta(0)$ vanishes, straightforwardly gives the logarithm of the functional determinant, and hence equal to the effective action.
Substituting into \eqref{W}, one can finally use the obtained expression to calculate the effective action corresponding to the theory with Lagrangian \eqref{masterL} and metric \eqref{sequence}.

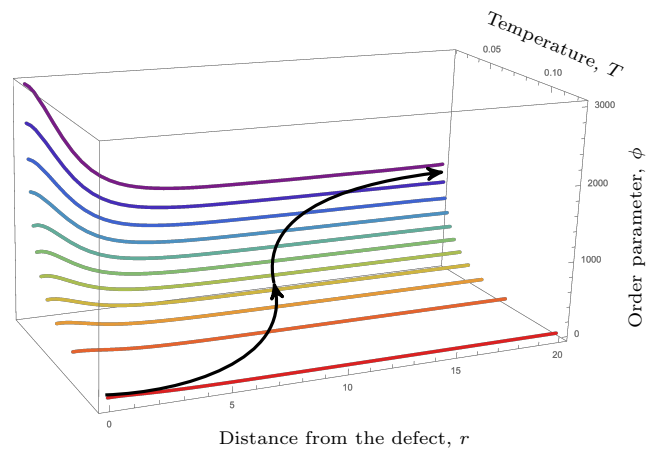
\begin{figure}[b!]
 \includegraphics[width=8cm]{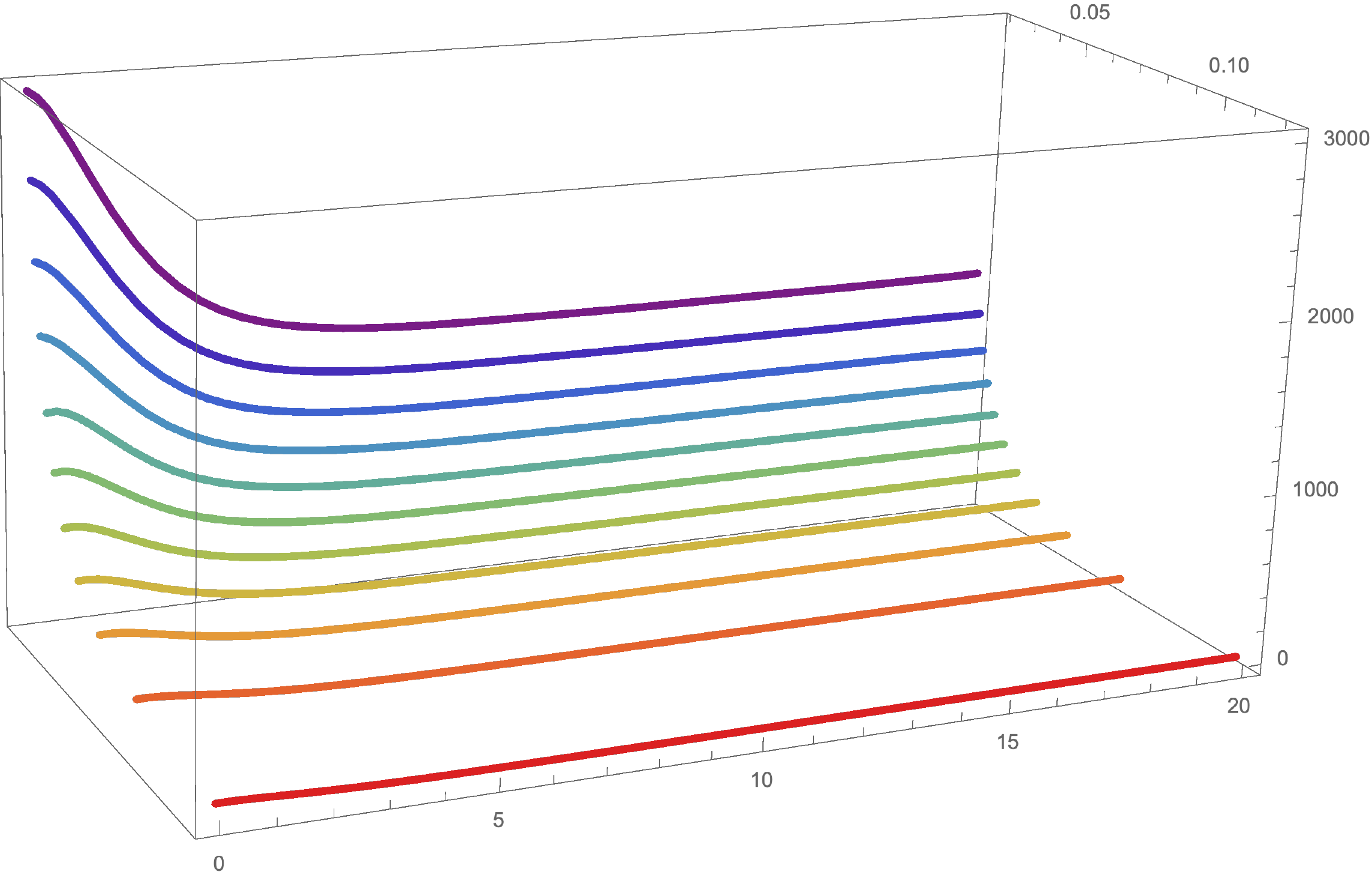}
       \put(-215,-12.5){\rotatebox{32}{\begin{tikzpicture}[->,>=stealth',auto,node distance=2.82 cm,very thick]  
       \node (1) {};
       \node (2) [right of=1] {};
              \node (3) [right of=2] {};
       \draw [->] (1) to [out=-30, in=250] (2.center);
             \draw [->] (2.center) to [out=70, in=155] (3);
%       \put(-210,-5){\rotatebox{32}{\begin{tikzpicture}[->,>=stealth',auto,node distance=2.81 cm,very thick]  
%       \node (1) {};
%       \node (2) [right of=1] {};
%              \node (3) [right of=2] {};
%       \draw [->] (1) to [out=-30, in=220] (2.center);
%             \draw [->] (2.center) to [out=40, in=155] (3);
\end{tikzpicture}}}
\put(-150,-5){\rotatebox{0}{{\scriptsize Distance from the defect, $r$}}}
\put(5,39){\rotatebox{90}{{\scriptsize Order parameter, $\phi$}}}
\put(-50,155){\rotatebox{-23}{{\scriptsize Temperature, $T$}}}
     \caption{Condensate profile as a function of the temperature, at zero chemical potential. Adjacent profiles are separated by $\delta\beta=4$. At a fixed temperature, the ordered phase is enhanced when approaching the defect. The black trajectory is an example of a condensate obtained modulating the temperature along with the distance from the defect. }\label{conduu}
\end{figure}

The effective action can now be numerically evaluated. 
%for any value of the chemical potential and temperature. 
As an example, Figs. \ref{pot} and \ref{conduu} show the results for zero chemical potential, positive curvature (cone) at different values of the temperature. The two figures in Fig. \ref{pot} show, respectively, the trends of the effective potential close to the apex of the regularised cone (left panel) and far away from the defect (right panel), as a function of the temperature. Note that, as expected, increasing the temperature corresponds to pushing the system toward a disordered phase. 
The numerical integration of the effective equations of motion finally gives an explicit profile for the order parameter, Fig. \ref{conduu}: the approach to the central defect enhances the symmetry breaking and the formation of the condensate, further confirmation of the effect discovered in \cite{Castro:2018iqt}, with temperature favouring the phase transition.

\section{Conclusion}

In quantum field theory, the Coleman-Weinberg mechanism \cite{colwei} explains how symmetries may spontaneously break as a result of quantum effects, predicting a phase transition (of the first order in scalar electrodynamics) from a broken to a restored symmetry phase as the mass is increased. Once the background geometry underlying the field theory is curved, the same mechanism leads to expecting a similar transition when the scalar curvature is increased. However, this is not the whole story, and when the topology of the background becomes non-trivial, new effects may take place. An example of the sort has been studied in Ref.~\cite{Castro:2018iqt}, with the background geometry emerging from the continuum limit of a deformed hexagonal lattice, with the deformation being due to the presence of a defect in the lattice tessellation. This ``kirigami-engineered'' background is quite special, since aside for altering the curvature it also modifies the boundary conditions that fields should obey when circulating the defect. 
%It can be shown (see Refs.~\cite{}) that t
These non-trivial boundary conditions can be mimicked by a synthetic gauge field localised near the apex and compete with the curvature in altering the ground state, inducing condensation close to the defect.

%it is indeed followed by the modification of the topological characteristics of the system, leading to some non-trivial boundary conditions. External conditions and structural properties affect the way symmetry breaking occurs in quantum theories defined on lattices. When the fixed, curved background is a discrete kirigami-engineered lattice, however, the change in the geometry does not come for free: it is indeed followed by the modification of the topological characteristics of the system, leading to some non-trivial boundary conditions. 

A relevant question is how additional external conditions change the picture. To address this problem, we have adapted to the present case the imaginary time formalism and zeta function regularisation techniques to arrive at a partially re-summed form of the effective action, whose minimisation has been carried out by numerical approximation. We have studied what happened when effects of finite temperature are included, resulting in a technically non-trivial modulation of the order parameter, as illustrated in Fig.~\ref{conduu}. Given a certain temperature, the condensate is always enhanced when approaching the defect, as a consequence of the kirigami effect. 

An intriguing possibility is to tune the temperature to obtain a more complicated pattern of the condensate as a function of the distance from the defect. The black curve of Fig.~\ref{conduu}, for example, shows the possible profile of a condensate obtained decreasing the temperature as stepping away from the defect. Here, the interesting aspect is in the value of the condensate, that increase with distance, rather than decreasing. The situation pictured in this way, however, is valid only as a first approximation. Temperature anisotropies, in fact, induce extra variation in the local density of the condensate and could make the effect of inhomogeneities (and hence of the contribution of the chemical potential part to the effective action) non-negligible.

%jointly with a smart re-summation of all the scalar curvature terms arising from the heat kernel expansion of the (squared) Dirac operator on the curved manifold, provides some general non-perturbative results to describe the behaviour of the order parameter emerging from symmetry breaking.These results should be relevant for applications to the physics of novel quantum materials. In the case of the bosonised Hubbard model, used to describe electron-electron interactions on a hexagonal lattice like graphene, the simultaneous breaking of the discrete $\mathbb{Z}_2$ and sub-lattice symmetries develops staggered magnetisation as an order parameter, which has been numerically shown to coagulate in the proximity of the defect apex, at low temperature and zero chemical potential.  

\section*{Acknowledgments}

We acknowledge the support of the Japanese Ministry of Education, Culture, Sports, and Science Program for the Strategic Research Foundation at Private Universities `Topological Science' (Grant No. S1511006) and of the Japanese Society for Promotion of Science (Grant No. P17763 and KAKENHI Grants No. 18K03626, No. 17F17763). We are grateful to Eduardo Castro, Pedro Ribeiro, Yuki Masaki and Lorenzo Fratino for useful discussions.

\end{document}